\documentclass{aa}

\usepackage{natbib}
\bibpunct{(}{)}{;}{a}{}{,} 
\usepackage{graphics}   
\usepackage{graphicx}   
\usepackage{epstopdf}
\usepackage{longtable}   
\usepackage{url}      
\usepackage{bm}        
\usepackage[varg]{txfonts}
\usepackage{pdflscape}
\usepackage{caption}
\usepackage{supertabular}
\usepackage{hyperref}   

\hypersetup{
    bookmarks=true,         
    unicode=false,          
    colorlinks=true,       
    linkcolor=blue,          
    citecolor=blue,        
    filecolor=blue,      
    urlcolor=blue           
}
\usepackage{color}

\begin{document}

\title{High surface magnetic field in red giants as a new signature of planet engulfment?} 

\author{Giovanni Privitera\inst{1,2}, Georges Meynet\inst{1}, Patrick Eggenberger\inst{1}, Cyril Georgy\inst{1}, Sylvia Ekstr$\ddot{\rm o}$m\inst{1}, Aline A. Vidotto\inst{3},
Michele Bianda\inst{2}, Eva Villaver\inst{4}, Asif ud-Doula\inst{5} 
}

 \authorrunning{Privitera et al.}

 \institute{Geneva Observatory, University of Geneva, Maillettes 51, CH-1290 Sauverny, Switzerland
 \and Istituto Ricerche Solari Locarno, Via Patocchi, 6605 Locarno-Monti, Switzerland
 \and School of Physics, Trinity College Dublin, The University of Dublin, Ireland
\and  Department of Theoretical Physics, Universidad Aut\'onoma de Madrid, M\'odulo 8, 28049 Madrid, Spain
\and  Penn State Worthington Scranton, Dunmore, PA 18512, USA
}

\date{Received /
Accepted}
\abstract  {Red giant stars may engulf planets. This may increase the rotation rate of their convective envelope, which could lead to strong dynamo-triggered magnetic fields.} 
{We explore the possibility of generating magnetic fields in red giants that have gone through the process of a planet engulfment. We compare them with similar models that evolve without any planets.
We discuss the impact of magnetic braking through stellar wind on the evolution of the surface velocity of the parent star.} 
{
By studying rotating stellar models with and without planets and an empirical relation between the Rossby number and the surface magnetic field, 
we deduced the evolution of the surface magnetic field along the red giant branch.
The effects of stellar wind magnetic braking were explored using a relation deduced from magnetohydrodynamics simulations.} 
{The stellar evolution model of a red giant with 1.7 M$_\odot$ without planet engulfment and with a time-averaged rotation velocity during the main sequence
equal to 100 km s$^{-1}$ shows
a surface magnetic field triggered by convection
that is stronger than 10 G only at the base of the red giant branch, that is, for gravities log $g > 3$. 
When a planet engulfment occurs, this magnetic field can also appear at much lower gravities,  that is,  at much higher luminosities along the
red giant branch. The engulfment of a 15 M$_J$ planet typically
produces a dynamo-triggered magnetic field stronger than 10 G for gravities between 
2.5 and 1.9.
We show that for reasonable magnetic braking laws for the
wind, the high surface velocity reached after a planet engulfment may be maintained sufficiently long to be observable.}
{High surface magnetic fields for  red giants in the upper part of the red giant branch are a strong indication of a planet engulfment or of an interaction with a companion. Our theory can be tested by observing
fast-rotating red giants such as HD31993, Tyc 0347-00762-1, Tyc 5904-00513-1, and Tyc 6094-01204-1 and by determining whether they show magnetic fields.}
\keywords{}

\maketitle

\titlerunning{Star-planet interactions}
\authorrunning{Privitera et al.}

\section{Introduction}

The tidal forces between star and planets change the planetary orbit and may sometimes produce the engulfment of the planet
by the star. A number of past studies 
\citep{soker84, siess99I, siess99II, villaver07, villaver09, nordhaus10, kunimoto11, villaver14} 
have shown that  the possible impact of an  engulfment, planetary or otherwise, leads to changes in the stellar 
luminosity and radii for a short  period of evolutionary time, to modifications of the surface abundances, in particular to
an increase in 
lithium abundance \citep{alexander67, fekel93, sandquist98, sandquist02, carlberg10, adamow12}, and to changes in the surface velocity \citep{carlberg09, carlberg14}.

Our team focused on this last consequence in two previous papers referred to here as Papers I and II \citep{priviter15,priviter16}. 
To explore this effect, we used stellar rotating models
where the rotation in the whole star can be followed consistently according to  the shellular theory by \citet{zahn92}
and affected by tidal interaction when an external convective zone appears \citep{zahn77, zahn89}.
Papers I and II  showed that
the observed surface velocities of some stars can only be explained by some interaction with a planet or a brown dwarf. 

In this letter we study to which extent such tidal interactions and engulfment processes may trigger
a magnetic field and also whether this {\it \textup{planet-induced}} magnetic field might be strong enough to slow down the stellar rotation
through the process of wind magnetic braking \citep{ud2002, ud2008}.
Section~\ref{sec:pom} briefly recalls the main ingredients of the models. The evolution of the Rossby number for different stellar models is discussed 
in Sect.~\ref{sec:ern}. Planet-induced magnetic fields and their possible consequences on the evolution of the stellar rotation are presented in Sect.~\ref{sec:pe_smf},
and conclusions are drawn in Sect.~\ref{sec:concl}.

\section{Physics of the models}\label{sec:pom}

To study the problem addressed here, we need to understand  stellar and planetary models whose evolutions are closely inter-linked through five different processes: (a) the evolution of the planetary orbit that accounts for the evolution of the star, (b) the changes
in 
angular momentum of the star that are due to changes in planetary orbit, (c) the physics of the engulfment, (d) the link between the engulfment and the generation of a surface magnetic field, and (e) the impact
of the generated magnetic field on the rotational evolution of the red giant star. The first three processes were extensively discussed in Papers I and II. Here we focus on the last two,
which involve magnetic fields. Red giants are characterized by a deep convective envelope in slow rotation, therefore tidal dissipation can be introduced by taking into account only the equilibrium tide \citep[see e.g.][]{ogilvie14,villaver14}. The impact of inertial waves in the convective envelope \citep[see e.g.][]{bolmont16} and dynamical tides in the radiative envelope of main-sequence models \citep{zahn75} can be safely neglected in the present computations because the convective envelope of red giants rotate slowly and the separation between the planet and the star (initial semi-major axis of 0.5 au) is relatively large.

As shown in Sect.~\ref{sec:pe_smf}, a planet engulfment may strongly accelerate the convective envelope of a red giant. 
When rotation is high enough, it may, through Coriolis acceleration, create differential rotation and helical turbulence in the convective zone, which is required for a dynamo process (the so-called $\alpha$ and $\omega$ effects, respectively) \citep[see e.g. Sect. 3.2.2 in][]{Charbonneau2013}.
To determine whether Coriolis acceleration is strong enough to create differential rotation, a dimensionless number called Rossby number can be defined as
\begin{equation}
{\rm Ro}\equiv P_{\rm rot}/t_{\rm tov}
,\end{equation}
where $P_{\rm rot}$ is the stellar spin period and $t_{\rm tov}$ the convective turn-over time. In general, ${\rm Ro}<1$ indicates the regime where the Coriolis acceleration strongly modifies convective flows and turbulence. We can readily compute the evolution of this number from our rotating stellar models using
the expression for the convective turn-over time given by Eq. (4) in \citet{rasio96}.

An exact theory linking rotation and magnetic field is still lacking, but an empirical relation showing that
the surface magnetic field ($B$) of red giants increases when the Rossby number decreases has recently been found for solar-type and low-mass stars \citep{Vidotto2014, Aurriere2015}.  
We here use the relation given by  \citet{Aurriere2015} for red-giant stars:
$\lg (B)=-0.85*\lg ({\rm Ro}) +0.51$.

Once a sufficiently strong surface magnetic field is generated, it may force the stellar wind to co-rotate with the star. This in turn may lead to a torque on the convective envelope. This is the essence of wind magnetic braking. The loss of angular momentum by this process is first estimated following \citet{ud2002} and \citet{ud2008},
\begin{equation}
{{\rm d}J\over {\rm d}t }={2 \over 3}\dot M \Omega R^2[0.29+(\eta_* + 0.25)^{1/4}]^2,
\label{ma}
\end{equation}
where $J$ is the angular momentum of the convective envelope,
$\dot M$ the mass-loss rate given by  \citet{rei75},
$\Omega$, the angular surface velocity, 
$R$ the stellar radius, 
and $\eta_*$ the magnetic confinement parameter \citep{ud2002} defined by 
$
\eta_* \equiv {B^2 R^2 / \dot M \upsilon_\infty},
\label{eta}
$
where $B$ is the surface magnetic field at the equator 
and $\upsilon_{\infty}$ the terminal wind velocity.
Winds of red giant stars are expected to be slow, 
 with observed terminal wind velocities of between 30 and 70 km s$^{-1}$
\citep[][]{Robinson1998}.
For our models here, we chose $\upsilon_\infty = 50$ km s$^{-1}$.
Except for $\upsilon_{\infty}$, the other quantities were taken from stellar evolution models.

Equation~(\ref{ma}) was deduced for hot massive stars from 2D MHD simulations of magnetic wind confinement models where a rotation-aligned dipole magnetic field is considered. In the case of red giants, one notable difference, however, is the ionization fraction of the wind, which is of course much lower for red giants than for the hot stars. The lower the ionization fraction, the lower the coupling between the wind and the magnetic field, probably, and thus the weaker the wind braking. To account for this effect, we computed the loss of angular momentum in our model (${\rm d}L/{\rm d}t$) using a parametric approach by introducing a wind braking efficiency factor $f \le 1$  so that  ${\rm d}L/{\rm d}t=f {\rm d}J/ {\rm d}t,$ where ${\rm d}J/ {\rm d}t$ is given by Eq.~(\ref{ma}).

Because Eq.~(\ref{ma}) was deduced for hot massive stars, the
question is whether it can be used for red giants with an external convective envelope. In principle, this can be the case if the formalism is not too sensitive to the wind-driving mechanism. To check this point, we also used the formalism of \citet{matt12}, which has been obtained for solar-type stars. The loss of angular momentum according to \citet{matt12} is then given by
\begin{equation}
\frac{{\rm d} J}{{\rm d}t}=\frac{K_{\rm 1}^{\rm 2}}{(2G)^{\rm m}}B^{\rm 4m}\dot{M}^{\rm 1-2m}\frac{R^{\rm 5m+2}}{M^{\rm m}}\frac{\Omega}{(K_{\rm 2}^{\rm 2}+0.5s_{\star}^{\rm 2})^{\rm m}}
\label{equa:ma_matt}
,\end{equation}
where $K_{\rm 1}=1.3$ and $K_{\rm 2}=0.0506$, $G$ is the gravitational constant, $m=0.2177$, $M$ is the mass of the star, and $s_{\star}=\Omega R^{3/2}(GM)^{-1/2}$ \citep[see][for more details]{matt12}. For Eq.~(\ref{ma}) an efficiency factor $f$ is used to change the braking efficiency.

\section{Evolution of the Rossby number}\label{sec:ern}

\begin{figure*}
\centering
\includegraphics[width=.50\textwidth]{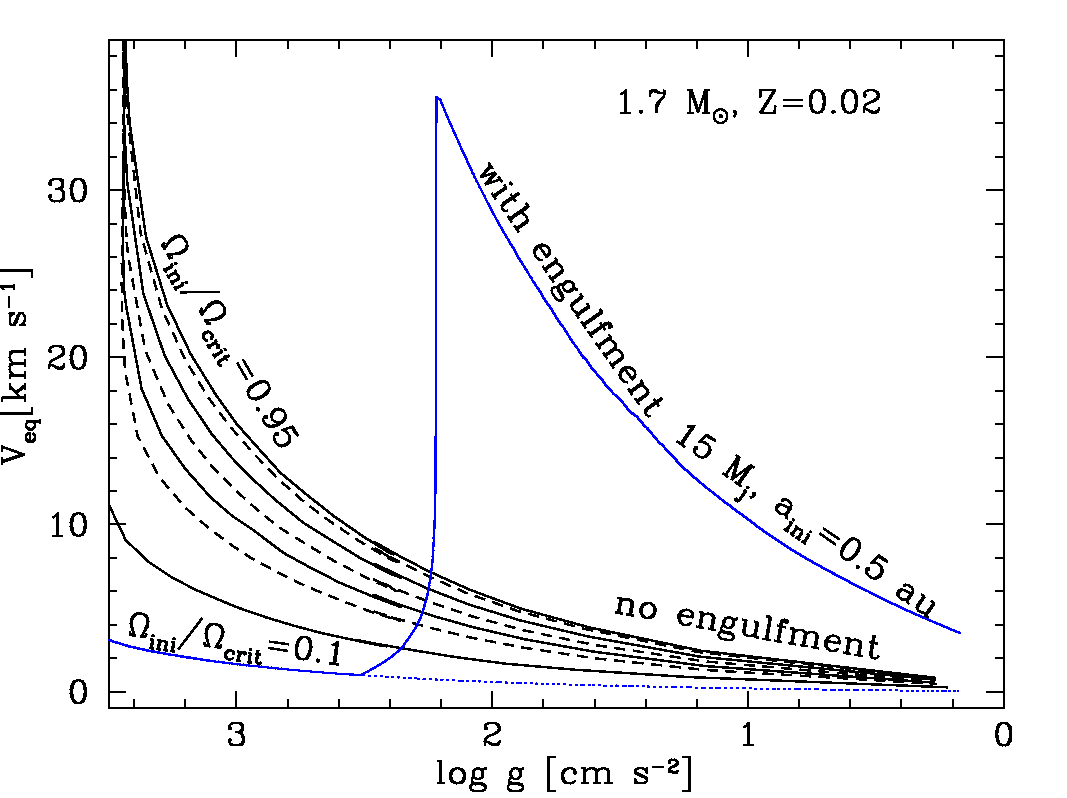}\includegraphics[width=.50\textwidth]{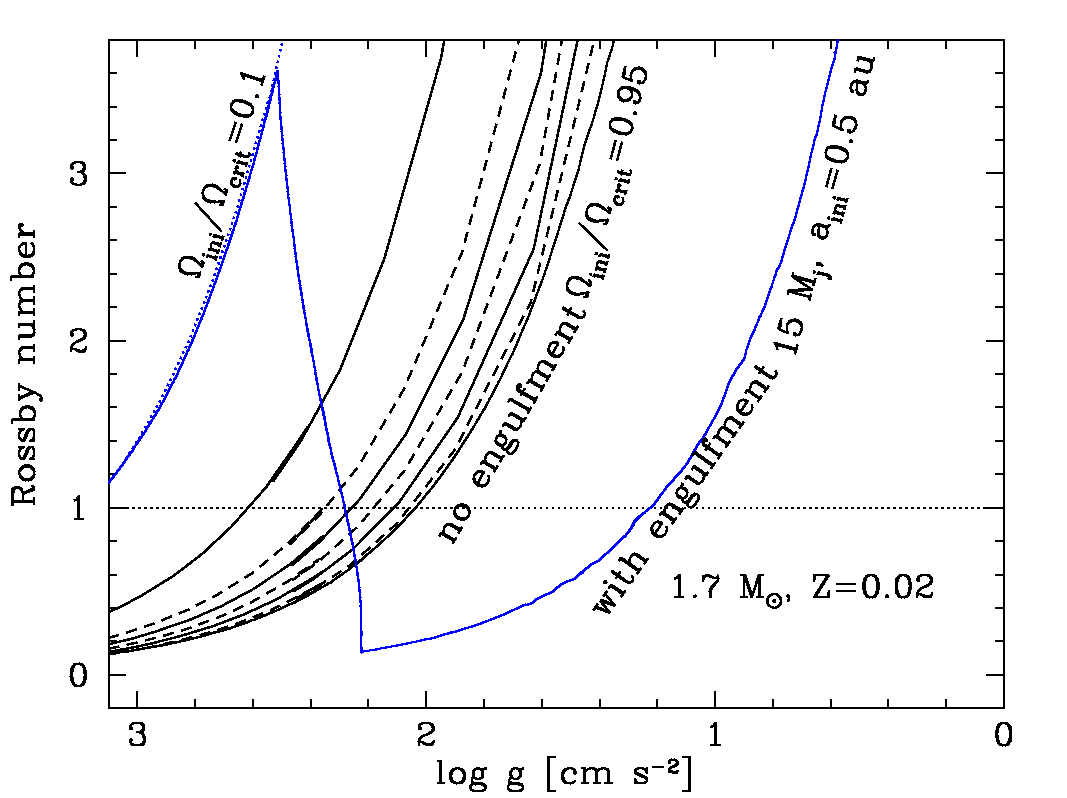}
\caption{{\it Left panel:} Predicted
evolution of the surface velocity as a function of the surface gravity for 1.7 M$_\odot$ stellar models along the red giant branch with and without planet engulfment.
Different initial rotations on the ZAMS are considered for models without engulfment: $\Omega_{\rm ini}/\Omega_{\rm crit}=$0.3, 0.5, 0.6, 0.7, 0.8, 0.9, and 0.95 (alternate black continuous and dashed lines). The time-averaged surface velocities
during the main-sequence phase of the models vary between about 30 and 240 km s$^{-1}$. The model with planet engulfment (blue continuous line, the dotted blue line is the model without engulfment) initially had $\Omega_{\rm ini}/\Omega_{\rm crit}=$0.1 and a 15 M$_J$ planet
orbiting at a distance equal to 0.5 au. The engulfment occurs when the surface gravity of the star is around 2.2.
{\it Right panel:} Evolution of the Rossby number as a function of the surface gravity for the same stellar models along the red giant branch with and without a planet engulfment.}
\label{vit}
\end{figure*}

Planet engulfment has a very significant impact on the surface rotation of red giants. We show this in the left panel of Fig.~\ref{vit}, where we plot the evolution of surface rotation velocities as a function of surface gravity for 1.7 M$_\odot$ stellar models with various initial rotation rates, as indicated in the caption. It then compares what happens when a 15 M$_J$ planet is engulfed (at a log $g$ of around 2.2)
by an initially slowly rotating 1.7 M$_\odot$ star. Clearly, the engulfment leads to a spin-up at a level that cannot be reached by single-star models regardless of  their initial rotation rates (Paper II). 
If a lower mass planet is considered, then this will change the time at which the engulfment occurs, shifting it to lower gravities, which will produce
a lower increase in the surface velocity (see Paper II). The same occurs when the mass of the planet is kept constant but the initial distance to the star is increased.

The right panel of Fig.~\ref{vit} shows the evolution of the Rossby number for the same models. Models without engulfment have  a phase where ${\rm Ro}<1$  at the base of the red giant branch when the star rotates sufficiently fast  for
${\rm Ro}$  to decrease below 1. More quantitatively, when we initially
consider the fastest rotating model,
${\rm Ro}<1$ only for surface gravities higher than about 2,
that is, {\it } at the base of the red giant branch. 
When an engulfment occurs, a quite different evolution of ${\rm Ro}$ is evident. 
At the time of engulfment, the Rossby number becomes suddenly smaller than 1. After the engulfment, the star continues to expand and the magnetic braking
causes the Rossby number to increase again. 

When we compare models with and without engulfment, the most striking difference is that
models with planet engulfment produce ${\rm Ro}<1$  {\it \textup{at lower
surface gravities}} (between 1.2 and 2.2), {\it } that is, at more advanced stages along the red giant branch star.  
In this range of surface gravities, no single-star model, regardless
of its initial rotation, predicts ${\rm Ro}<1. $ This suggests that such a model can hardly produce an observable magnetic field.

\section{Planet engulfment and surface magnetic field}\label{sec:pe_smf}

\begin{figure*}
\centering
\includegraphics[width=.50\textwidth]{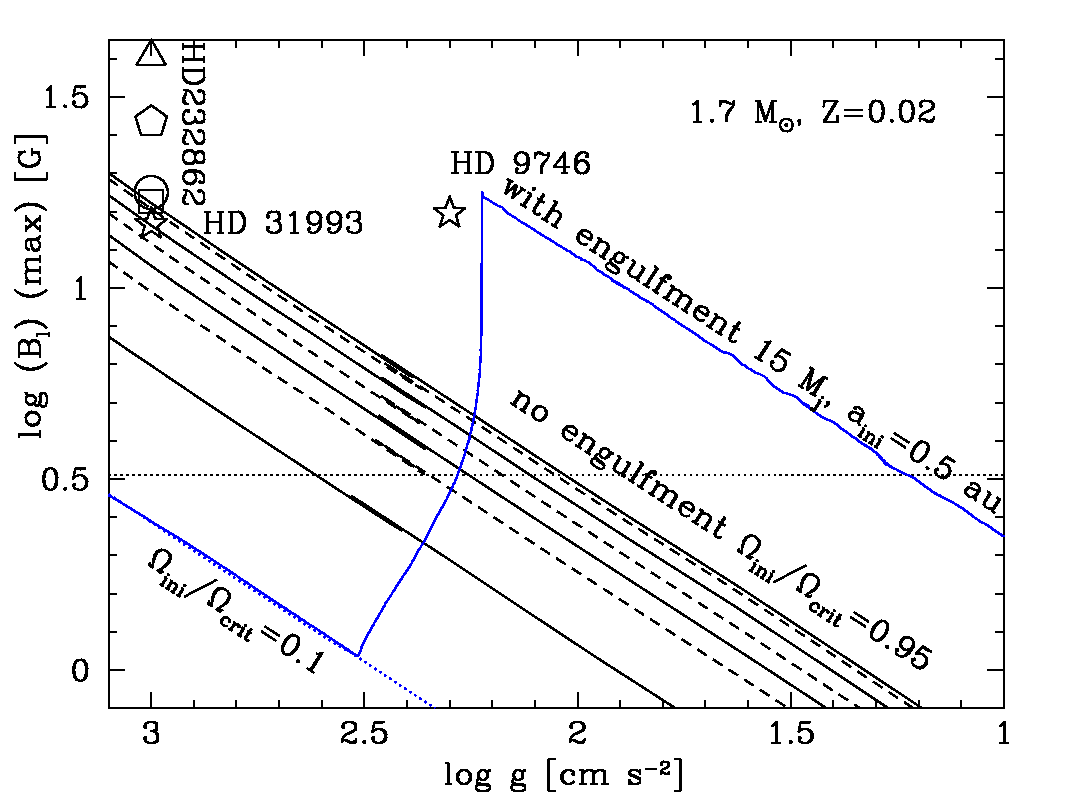}\includegraphics[width=.50\textwidth]{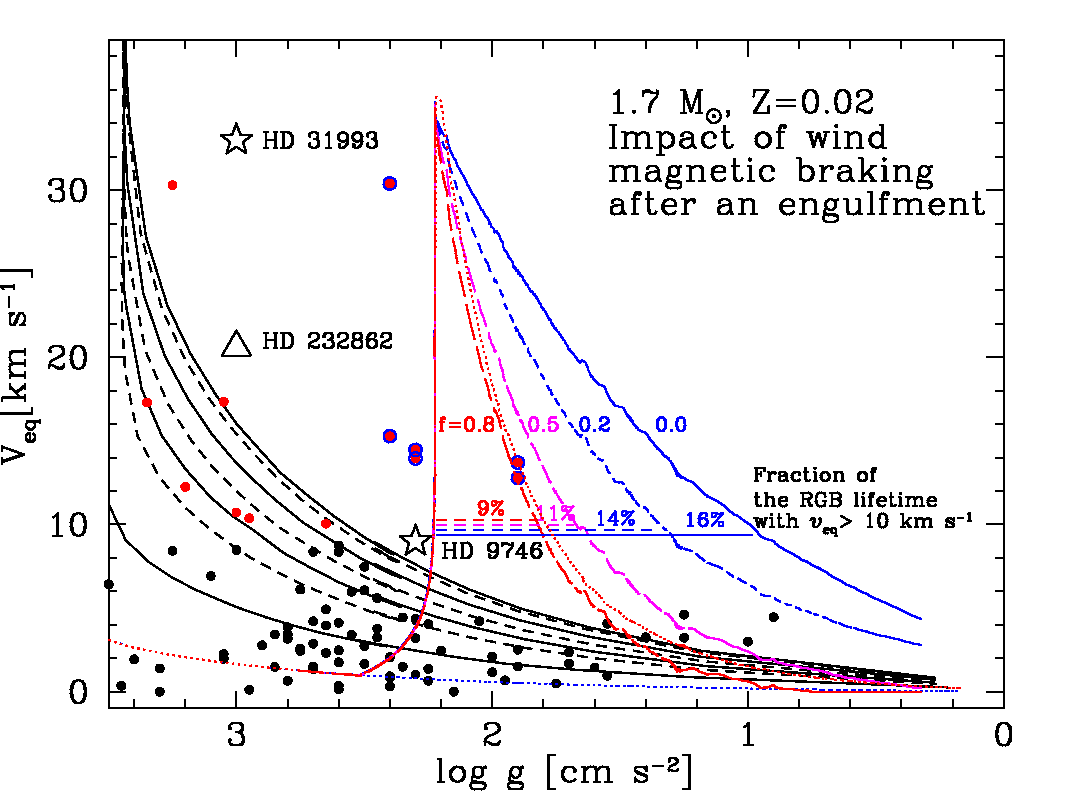}
\caption{{\it Left panel:} Predicted
evolution of the surface magnetic field as a function of the surface gravity for 1.7 M$_\odot$ stellar models along the red giant branch with and without planet engulfment.
The models are the same as  in Fig.~\ref{vit}. Above the horizontal dashed line, ${\rm Ro}<1$. The two empty stars are observations from \citet{Aurriere2015}. The other points correspond to the star HD232862 observed on four consecutive days by \citet{Lebre2009} (the triangle is the first observation, then the square, pentagon, and the circle).
{\it Right panel:} Evolution of the surface rotation as a function of the surface gravity for the same models as shown in Fig.~\ref{vit}, with in additional cases where magnetic braking laws with various efficiencies have
been accounted for after the engulfment. The dashed blue, magenta, and red curves correspond to values of $f$ equal to  0.2, 0.5, and 0.8 using Eq.~(\ref{ma}) (see text). The red dotted line is obtained using Eq.~(\ref{equa:ma_matt}) with $f=0.5$. The continuous blue curve corresponds to the case $f=0$.
We have indicated how the duration of the period during which the surface velocity is above 10 km s$^{-1}$ varies as a function of the strength of the magnetic braking.
The same three observations as those indicated in the left panel are shown. The dots show the observations by \citet{Carlberg2012}, the black  circles show the stars with a $\upsilon \sin i < 10$ km s$^{-1}$, the filled red- and blue-circled magenta points
have $\upsilon \sin i > 10$ km s$^{-1}$. The blue-circled magenta points correspond to stars (HD31993, Tyc 0347-00762-1, Tyc 5904-00513-1,
and Tyc 6094-01204-1) whose surface velocity cannot be explained by any reasonable model for single stars 
(Paper II). 
}
\label{field}
\end{figure*}

We now study the magnetic field generation for our models. The
left panel of Fig.~\ref{field}
shows the predicted evolution of the surface magnetic field as a function of the surface gravity for 1.7 M$_\odot$ stellar models  with and without planet engulfment.
Surface magnetic fields of up to a few tens of Gauss could be triggered
by an engulfment at surface gravities where such fields are  otherwise not expected.
Even for the initially faster rotating stars that evolved in isolation, no significant magnetic field is expected  for
log $g <2$. Thus, a strong magnetic field at low gravities together with a high surface velocities are strong signs of
a past engulfment. Of course, a planet with too low a mass or a planet initially orbiting the star at too large a distance will only lead to a weak magnetic field that will not be observable.

The question here might be asked whether the strong magnetic field linked to the fast rotation 
might prevent the star from keeping trace of the\textup{ {\it \textup{past engulfment high rotation rate}}} for a sufficiently long time to be an
observable feature. 
When our 1.7 M$_\odot$ reaches its peak surface velocity at the time of engulfment,
the surface magnetic field is around 20 G. 
With such a strong field, the shortest wind magnetic braking timescale (obtained using $f$=1) is  20 Myr. 
Since the duration from the engulfment to the tip of the red giant branch is about 34 Myr, 
wind magnetic braking is expected to significantly affect the evolution of the surface velocity (of course,
a value of $f$=1 is most likely an overestimate caused by the low ionization fraction of the wind). 

In the right panel of Fig.~\ref{field} we show the results of various braking laws of the type given by Eq.~(\ref{ma}), with different wind braking efficiencies  $f$, as indicated in the caption. 
The wind magnetic braking was applied only to the case with engulfment.

The magnetic braking does not change the maximum velocity reached by the engulfment. The rise is too short for this braking to have any effect.
After the engulfment, the stronger the braking, the more rapidly the decrease in surface velocity. Without magnetic braking, the star retains
a surface velocity above 10 km~s$^{-1}$ for 29 Myr. When $f$ equals 0.2, 0.5, or 0.8, this duration reduces to 25, 20, and 16 Myr, respectively.
Compared to the whole red giant branch lifetime, which is about 181 Myr, these values correspond to fractions equal to 16\% ($f$=0),  14\% ($f$=0.2), 11\% ($f$=0.5),
and 9\% ($f$=0.8), thus still quite substantial values. This
means that even though the wind magnetic braking counteracts the effect of the engulfment on the surface velocities,
planet engulfment remains a strong candidate for explaining the observed fast-rotating red giants.

All these results were obtained by using the magnetic braking formalism of \citet{ud2002} and \citet{ud2008} (Eq.~(\ref{ma})). A model computed with the formalism of \citet{matt12} (Eq.~(\ref{equa:ma_matt})) is shown by the red dotted line in the right panel of Fig.~\ref{field}. A value of $f$ equal to 0.5 was used in this case. The two braking formalisms lead to very similar results (only a slight increase of the global efficiency of the braking is seen when Eq.~(\ref{equa:ma_matt}) is used). This can be understood by comparing for instance the expressions for the Alfv\'en radius in the two cases (Eq.~(19) in \citet{ud2008} and Eq.~(6) in \citet{matt12}): for red giants, the term related to the magnetic confinement parameter dominates and the dependence of the Alfv\'en radius on this term is nearly identical in both formalisms (a value of $m$ of 0.2177 for \citet{matt12} and 0.25 for \citet{ud2008}).

Wind magnetic braking would also affect the evolution of the surface velocities of our models with no engulfment. These stars, as shown above, may also develop a surface magnetic field at the base of the red giant branch
(see right panel of Fig.~\ref{vit} and the left panel of Fig.~\ref{field}), which, at its turn, may drive some wind magnetic braking.  This would
shift the curves corresponding to {\it \textup{no engulfment}} in the surface velocity downward compared to the surface gravity plots (see the left panel of Fig.~\ref{vit} and the right panel of Fig.~\ref{field}),
but would not  much affect what occurs during the engulfment.  For the case considered here,  the quantity of angular momentum added by the planet
is so large that the initial surface rotation of the star has little influence.

Wind magnetic braking during the red giant phase would change the interpretation of the observed surface rotations of red giants stars.
Wind magnetic braking would shift to lower values the maximum surface velocity that can be  reached at a given surface gravity by an initial mass model that evolves without interaction (tides or engulfment).
The maximum  surface velocity is obtained by considering the highest possible initial rotation on the ZAMS and assuming solid-body rotation.
Any stars presenting higher surface rotations than this limit are very strong candidates for having experienced planet engulfment (see more in Paper II). 
Since wind magnetic braking would lower this limit, it would allow more observed stars to lie above it and thus, as indicated above, would enlarge the size
of stars whose surface rotation needs an interaction to be explained. A too efficient magnetic braking would probably increase the
number of candidates that would have had to experience such interactions by too much.
This point will be discussed in a more extended future work.

Although it is beyond the scope of the present work to make detailed comparisons with observations, we discuss below three stars
that are interesting candidates for having engulfed a planet, and that 
have masses around 1.7$\pm0.3$M$_\odot$ {\it } , that is, they
are similar to the models discussed here.
The positions of these stars in the magnetic field versus log $g$ and in the surface velocity versus log $g$ plots are
indicated in Fig.~\ref{field}. The daily variation of the magnetic field of HD232862
may be due to the change of the viewing angle when the star rotates.
These three stars have surface magnetic fields and rotations that are high enough for being
good candidates for a planet engulfment.  Two of them (HD 31993 and 232862) might also be explained by assuming
initially very fast-rotating stars, but the probability of such extreme initial conditions is quite low so that
an interaction remains a more reasonable explanation for the
currently observed properties.
For these two stars, the engulfment of a 15 M$_J$ planet that
began orbiting the star at 0.5 au occurs at a gravity that is
too low to explain their properties. This would indicate a higher mass planet or a shorter initial distance between planet and star.
The magnetic field of HD9746 may well be explained by the case of engulfment shown in the figures. 
Its $\upsilon \sin i$ is equal to 9 km s$^{-1}$ \citep{Bal2000}. Our models
would predict a value $\upsilon$  around 30 km s$^{-1}$.
Planet-induced magnetic field theory might be tested by checking whether
those stars whose high surface velocities cannot be explained by stellar models evolving in isolation truly present measurable magnetic fields.  
Typically, stars such as HD31993, Tyc 0347-00762-1, Tyc 5904-00513-1, and Tyc 6094-01204-1 (the blue-circled magenta points in Fig.~2) are 
interesting candidates.

\section{Conclusion}\label{sec:concl}

A planet engulfment can increase stellar rotation to such an extent that an observable surface magnetic field can develop.
Planet engulfment might explain the strong magnetic field in the upper part of the red giant branch, where
stars without interaction rotate too slowly to allow a dynamo {\textbf to create such strong fields}. Some
good candidate stars show both high surface rotation rates and high surface magnetic fields
that are compatible with an engulfment. Likewise, some fast-rotating red giants may be found to show some
surface magnetic fields if our present models are correct (for instance HD31993, Tyc 0347-00762-1, Tyc 5904-00513-1, and Tyc 6094-01204-1).

The planet-induced magnetic fields are strong enough to slow the star down by wind magnetic braking. This will reduce the time
that the surface velocity of a star can be maintained above a given limit. However, for moderate couplings, the wind magnetic braking effect does not disrupt
the production of observable ({\it i.e.} sufficiently long-lasting) fast-rotating red giants by planet engulfment. This seems compatible with recent results that suggested a low efficiency of surface magnetic braking for stars more evolved than the Sun \citep{vansaders16}.

The increased stellar rotation and magnetic field generation linked to planet engulfment can in principle operate on the wind
and might increase the mass loss. 
Whether this mechanism  can explain the high mass-loss rates needed to form hot subdwarf stars  that are singles
is an interesting question that needs to be explored in the future.

\begin{acknowledgements} 
We would like to thank St\'ephane Mathis for his valuable comments and suggestions. The project has been supported by Swiss National Science Foundation grants 200021-138016,  200020-160119 and 200020-15710. E.V. acknowledges support from the Spanish Ministerio de Economía y Competitividad under grant AYA2014-55840P.
AuD acknowledges support by NASA through Chandra Award numbers GO5-16005X, AR6-17002C and  G06-17007B issued by the Chandra X-ray Observatory Center which is operated by the Smithsonian Astrophysical Observatory for and behalf of NASA under contract NAS8- 03060.
\end{acknowledgements}

\bibliographystyle{aa} 
\bibliography{biblio.bib} 


\end{document}